\makeatletter
\newcommand{\dontusepackage}[2][]{%
  \@namedef{ver@#2.sty}{9999/12/31}%
  \@namedef{opt@#2.sty}{#1}}
\makeatother
\dontusepackage{subfigure}

\documentclass[]{article}

\usepackage{lmodern}
\usepackage{amssymb,amsmath}
\usepackage{ifxetex,ifluatex}
\usepackage[usenames,dvipsnames]{color}
\usepackage{fixltx2e} % provides \textsubscript
\ifnum 0\ifxetex 1\fi\ifluatex 1\fi=0 % if pdftex
  \usepackage[T1]{fontenc}
  \usepackage[utf8]{inputenc}
\else % if luatex or xelatex
  \ifxetex
    \usepackage{mathspec}
    \usepackage{xltxtra,xunicode}
  \else
    \usepackage{fontspec}
  \fi
  \defaultfontfeatures{Mapping=tex-text,Scale=MatchLowercase}
  
\fi
\IfFileExists{upquote.sty}{\usepackage{upquote}}{}
\IfFileExists{microtype.sty}{%
\usepackage{microtype}
\UseMicrotypeSet[protrusion]{basicmath} % disable protrusion for tt fonts
}{}
\usepackage[margin=1.0in,bottom=1.5in]{geometry}
\usepackage[]{natbib}
\usepackage{listings}
\usepackage{graphicx}
\makeatletter
\def\maxwidth{\ifdim\Gin@nat@width>\linewidth\linewidth\else\Gin@nat@width\fi}
\def\maxheight{\ifdim\Gin@nat@height>\textheight\textheight\else\Gin@nat@height\fi}
\makeatother
\setkeys{Gin}{width=\maxwidth,height=\maxheight,keepaspectratio}
\usepackage{caption}
\usepackage{float}
\usepackage{subfig}
\ifxetex
  \usepackage[setpagesize=false, % page size defined by xetex
              unicode=false, % unicode breaks when used with xetex
              xetex]{hyperref}
\else
  \usepackage[unicode=true]{hyperref}
\fi
\hypersetup{breaklinks=true,
            bookmarks=true,
            pdfauthor={},
            pdftitle={A practical workflow for land seismic wavefield recovery with weighted matrix factorization},
            colorlinks=true,
            citecolor=black,
            urlcolor=blue,
            linkcolor=black,
            pdfborder={0 0 0}}
\def\vector#1{\mathbf{#1}}

\DeclareMathOperator\minim{\hbox{minimize}}

\usepackage{algorithm2e}
\def\minim{\mathop{\hbox{minimize}}}
\title{A practical workflow for land seismic wavefield recovery with weighted
matrix factorization}
\author{Yijun Zhang and Felix J. Herrmann,\\Department of Electrical \& Computer
Engineering, Georgia Institute of Technology\\}
\date{}

\begin{document}
\maketitle
\begin{abstract}
While wavefield reconstruction through weighted low-rank matrix
factorizations has been shown to perform well on marine data,
out-of-the-box application of this technology to land data is hampered
by ground roll. The presence of these strong surface waves tends to
dominate the reconstruction at the expense of the weaker body waves.
Because ground roll is slow, it also suffers more from aliasing. To
overcome these challenges, we introduce a practical workflow where the
ground roll and body wave components are recovered separately and
combined. We test the proposed approach blindly on a subset of the $3$D
SEAM Barrett dataset. With our technique, we recover densely sampled
data from $25$ percent randomly subsampled receivers. Independent
comparisons on a single shot demonstrate significant improvements
achievable with the presented workflow.
\end{abstract}

\vspace*{-0.45cm}

\section{Introduction}\label{introduction}

One of the critical phases in the early stages of oil and gas
exploration is seismic data acquisition. Inspired by relatively recent
developments encouraged by the field of Compressive Sensing
\citep{candes2006robust}, seismic data is increasingly collected
randomly along the spatial coordinates to shorten the acquisition time
and to reduce cost. While random sampling improves acquisition
productivity \citep{mosher2014increasing}, it does shift the burden from
field acquisition to data processing \citep{chiu20193d} since fully
sampled seismic data is a prerequisite to subsequent steps such as
multiple removal and migration.

Wavefield recovery is one of the key steps to reconstruct fully seismic
data from subsampled data. Recovery methods based on wavefield
reconstruction that exploits data sparsity in different transform
domains, such as wavelet \citep{villasenor1996seismic}, Fourier
\citep{sacchi1998interpolation}, and curvelet \citep{herrmann2008non},
have been proposed. More recently, several seismic studies have
investigated wavefield recovery via low-rank matrix factorizations
\citep{kumar2015efficient}, which are relatively simple and
computationally cheap. The general idea of these methods is to exploit
low-rank structure of fully sampled frequency slices when they are
organized in a matrix. \citet{oropeza2011simultaneous} and
\citet{kumar2015efficient} showed that the presence of noise or missing
traces increases the rank of these matrices, and they used this property
to recover the fully sampled frequency slices via low-rank matrix
factorization.

While the low-rank matrix factorization method has had some success,
especially for low to midrange frequencies, it struggles to recover high
frequency slices, which require higher ranks because they cannot be
accurately approximated by low-rank factorizations. To solve this
problem, \citet{aravkin2014fast}, \citet{eftekhari2018weighted}, and
\citet{zhang2019high} used the wavefield recovery via weighted matrix
factorization to reconstruct seismic data by introducing matrix weights
defined in terms of factorizations at neighboring frequencies that live
in close-by subspaces. By moving the matrix weights from the constraint
to the data-misfit term, \citet{zhang2020wavefield} proposed a
computationally more efficient scheme capable of handling high
frequencies.

Even though this weighted approach has had success, there remains the
challenge that land seismic data contains ground roll, which because of
its strong amplitude and high spatial frequency content is known to
\citep{liu1999ground} degrade the wavefield reconstructions based on
promoting structure whether this is sparsity or low rank. The reason for
this possible degradation is two-fold. First, groud roll corresponds to
Rayleigh-type surface waves, which are slow and for this reason often
aliased. Second, ground roll has strong amplitudes, which causes the
reconstruction to focus on the ground roll at the expense of
reconstructing the low-amplitude body waves. While ground roll is
typically dominant at the low temporal frequencies, its separation from
body waves is complicated by the fact that it is spatially aliased. By
reconstructing the wavefield to a fine grid, where the ground roll is no
longer aliased, we allow for a separation of ground roll and body waves
using $f-k$ filtering \citep{yilmaz2001seismic} or Radon domain
techniques \citep{trad2003latest}. During this talk, we present a
practical workflow aimed at removing the complications of carrying out
wavefield reconstruction on land data dominated by ground roll.

We organize this expanded abstract as follows. First, we discuss the
seismic wavefield reconstruction via weighted matrix factorization.
Next, we discuss the impact of ground roll. And then, we introduce our
proposed practical workflow step by step. We conclude by demonstrating
our approach on synthetic $3$D data simulated from the Barrett model and
show improved recovery quality compared to the conventional workflow.

\vspace*{-0.45cm}

\section{Reconstruction with weighted matrix
factorizations}\label{reconstruction-with-weighted-matrix-factorizations}

In \citet{aravkin2014fast}, \citet{eftekhari2018weighted} and
\citet{zhang2019high}, the authors proposed a wavefield recovery via
weighted matrix factorization. These factorizations are carried out on
data organized in monochromatic frequency slices and involve the
following optimization problem:
\begin{equation}
\begin{aligned}
& \minim_{\vector{X}_i} \quad \|\vector{Q}\vector{X}_i\vector{W}\|_*\\
& \text{subject to} \quad \|\mathcal{A}(\vector{X}_i) - \vector{B}_i\|_F \leq \eta.
\end{aligned}
\label{eqWlrmf}
\end{equation}
 In this expression, the symbol $\|\cdot\|_*$ represents the nuclear
norm, given by the sum of the singular values, and $\|\cdot\|_F$ denotes
the Frobenius norm, the energy of the matrix entries. The matrix
$\vector{X}_i$, for $i \in [1, \ldots , N_f]$, represents a fully
sampled monochromatic frequency slice at the $i\mathrm{th}$ frequency.
$N_f$ corresponds to the number of frequencies, and the matrix
$\mathcal{A}(\cdot)$ represents a mask operator used to subsample the
fully sampled frequency slice. The matrix $\vector{B}_i$ represents the
observed input data with missing traces at the $i\mathrm{th}$ frequency.
The misfit tolerance $\eta$ depends on the noise level in the observed
data.

To exploit the fact that seismic data exhibits low-rank behavior in the
so-called non-canonical organization \citep{kumar2015efficient}, we
matricize the frequency slices in the source-x receiver-x
organization---i.e., the to-be-recovered monochromatic data is
represented by the matrix
$\vector{X}_i \in \mathbb{C}^{(N_{sx}\times N_{rx}) \times (N_{sy}\times N_{ry})}$
where $N_{sx}$, $N_{sy}$ are the number of sources along the $x$ and $y$
coordinates, respectively. $N_{rx}$, $N_{ry}$ are the corresponding
numbers of receivers. The
$\{\vector{Q},\vector{W}\} \in \mathbb{C}^{(N_{sx}\times N_{rx}) \times (N_{sx}\times N_{rx})}$
are the weighting matrices, which include information on the subspaces
of a neighboring factorization as we reconstruct the wavefield from
low-to-high frequencies \citep{zhang2019high}. These weighting matrices
are given by
\begin{equation}
 \vector{Q} = {w}_{1}\vector{U}\vector{U}^H + \vector{U}^\perp \vector{U}^{{\perp}{H}}
\label{eqprojl}
\end{equation}
 and
\begin{equation}
\vector{W} = {w}_{2}\vector{V}\vector{V}^H + \vector{V}^\perp \vector{V}^{{\perp}{H}}. 
\label{eqprojr}
\end{equation}
 In these expressions, the symbol $^{H}$ denotes the Hermitian
transpose. The projection matrices
$\vector{U} \in \mathbb{C}^{(N_{sx} \times N_{rx}) \times r}$,
$\vector{V}\in \mathbb{C}^{(N_{sy}\times N_{ry}) \times r}$ contain rank
$r$ column and row subspaces that derive from neighboring (lower)
frequencies. The matrices $\vector{U}^\perp$, $\vector{V}^\perp$ are the
orthogonal complements of $\vector{U}$, $\vector{V}$. Because
factorizations of neighboring (lower) frequencies share information with
the current frequency slice, they can serve as prior information aiding
the wavefield recovery. The scalars ${w}_{1} \in \left(0,1 \right]$ and
${w}_{2} \in \left(0,1 \right]$ quantify the similarity between prior
information and the current to-be-recovered frequency slice. Small
values for these scalars indicate that we have more confidence in the
prior information.

As shown in \citep{zhang2019high}, considerable improvements can be made
during the recovery when reliable prior information is available.
However, including weighting matrices in the nuclear norm objective
function complicates the optimization making the minimization in
equation~\ref{eqWlrmf} computationally more expensive. To avoid this
issue, we follow \citet{zhang2020wavefield} and rewrite
equation~\ref{eqWlrmf} into
\begin{equation}
\begin{aligned}
& \minim_{\vector{\tilde {X}}_i} \quad \|\vector{\tilde {X}}_i\|_* \\
& \text{subject to} \quad \|\mathcal{A}({\vector{Q}^{-1}}\vector{\tilde{X}}_i{\vector{W}^{-1}}) - \vector{B}_i\|_F \leq \eta .
\end{aligned}
\label{eqWlrmfv2}
\end{equation}
 To arrive at this formulation, we replace the optimization variable
with $\vector{\tilde{X}}_{i} = \vector{Q}\vector{X}_i\vector{W}$. After
solving equation~\ref{eqWlrmfv2}, the original solution $\vector{X}_i$
can be recovered by
$\vector{X}_i=\vector{Q}^{-1}\vector{\tilde{X}}_i\vector{W}^{-1}$.
Mathematically, equations~\ref{eqWlrmf} and equation~\ref{eqWlrmfv2} are
equivalent except that the solution of the second formulation is easier
to compute by moving the weighting matrices to the data misfit
constraint.

To prevent computationally expensive singular value decompositions
(SVDs) part of the nuclear norm computations, we write
equation~\ref{eqWlrmfv2} in the following factored form:
\begin{equation}
\begin{aligned}
& \minim_{\vector{\tilde L}_i, \vector{\tilde R}_i} \quad \frac{1}{2} {\left\| \begin{bmatrix} \vector{\tilde L}_i \\ \vector{\tilde R}_i \end{bmatrix} \right\|}_F^2 \\
& \text{subject to} \quad \|\mathcal{A}{({\vector{Q}^{-1}} \vector{\tilde L}_i \vector{\tilde R}_i^{H} {\vector{W}^{-1}})} - \vector{B}_i\|_{F} \leq \eta.
\end{aligned}
\label{eqwlrf}
\end{equation}
 In this expression, the
$\vector{\tilde L}_i \in \mathbb{C}^{(N_{sx}\times N_{rx}) \times r}$
and
$\vector{\tilde R}_i \in \mathbb{C}^{(N_{sy}\times N_{ry}) \times r}$
represent the low-rank factorization of $\vector{\tilde X}_i$ with rank
$r \ll \min(N_{sx}\times N_{rx}, N_{sy}\times N_{ry})$
\citep{zhang2020wavefield}.

While wavefield recovery based on weighted matrix factorization has been
applied successfully (see e.g., \citet{zhang2019high} and
\citet{zhang2020wavefield}), its performance is challenged by data that
contains strong-amplitude aliased ground roll. Because of its
large-amplitude, ground roll dominates the reconstruction at the expense
of body waves that are of prime interest. In the next section, we will
introduce a practical workflow addressing this challenge.

\vspace*{-0.45cm}

\section{Impact of ground roll}\label{impact-of-ground-roll}

Acquisition and processing of land data are often challenged because it
is contaminated by strong ground roll. Because ground roll is slow, it
is often spatially aliased, complicating subsequent processing efforts
to remove this coherent noise component with $f-k$ or Radon filtering
\citep[\citet{trad2003latest}]{yilmaz2001seismic}. Unfortunately, it is
financially unfeasible to decrease the periodic receiver sampling
interval to avoid aliasing \citep{bahia2020ground}, and we have to
resort to alternative randomized acquisition methodologies
\citep[\citet{kumar2015efficient}]{mosher2014increasing} that are in
principle conducive to in silico unaliased wavefield reconstruction.
While this has proven to work, the presence of strong-amplitude ground
roll complicates wavefield reconstruction.

\begin{figure}
\centering
\includegraphics[width=0.500\hsize]{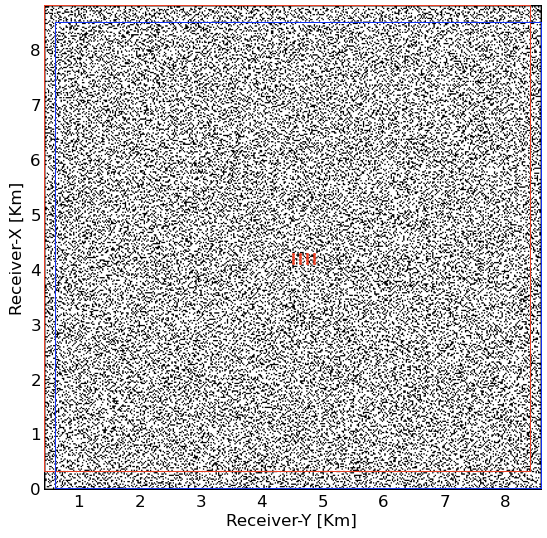}
\caption{Acquisition geometry for the recovered patch. Black $\cdot$'s
represent receiver locations, and red $\cdot$'s in the middle represent
the source locations. The red rectangle is the receiver aperture for the
top left source and the blue rectangle is the receiver aperture for the
bottom right source.}\label{geometry}
\end{figure}

To investigate this issue, we collaborate with Klaas Koster from
Occidental to conduct a blind study where we were provided with a subset
consisting of $21$ source lines, extracted from the synthetic $3$D SEAM
Barrett dataset \citep[\citet{tan2019seam}]{van2019sceptic}. This
dataset is designed to benchmark land data processing. As part of this
blind study, with the acquisition geometry plotted in
Figure~\ref{geometry}, we receive 3D shot records that are randomly
subsampled along the receivers. The $8\times 8$km receiver aperture is
moving with the source location, which means that between neighboring
shots randomly sampled receivers are mostly shared while some drop-off
and others are added (cf.~red and blue rectangle in
Figure~\ref{geometry}). Approximately 75 percent of receiver positions
are missing from the regular densely sampled periodic grid of $12.5$m,
yielding an effective average sample interval of $50$m, which is well
below Nyquist. The data consists of $667$ time samples with a sample
interval of $0.006\, \mathrm{s}$. The shots are sampled periodically
with a sample interval of $25$m in the shot-line direction and $100$m in
the perpendicular direction.

\begin{figure}
\centering
\subfloat[\label{UnweightedTime}]{\includegraphics[width=0.500\hsize]{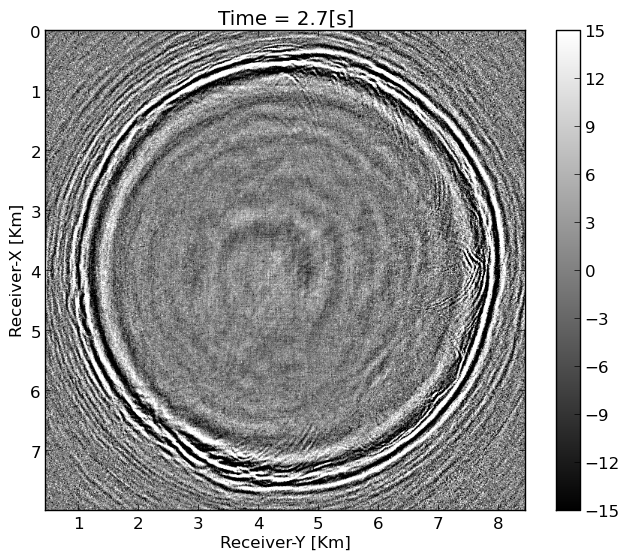}}
\subfloat[\label{UnweightedRX}]{\includegraphics[width=0.500\hsize]{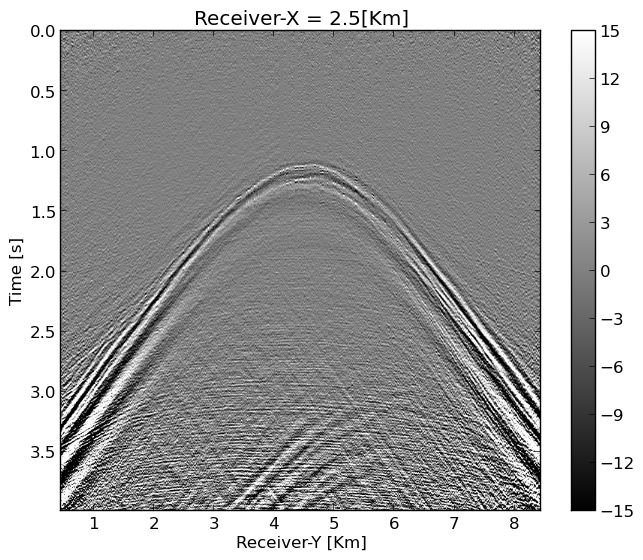}}
\caption{One common-shot record obtained by factorized wavefield
reconstruction. \emph{(a)} Time slice at $2.7\, \mathrm{s}$. \emph{(b)}
Shot record along the $y$ direction.}\label{OriginalData}
\end{figure}

To illustrate, the effects of the strong ground roll on our factorized
low-rank wavefield reconstruction scheme, we recover a patch of
$4\times 4$ shots with on the dense periodic receiver grid of $641$
receivers in each direction sampled at $12.5$m. This recovery
corresponds to solving a total of $384$ monochromatic matrix
factorization problems involving data volumes of
$667 \times 641 \times 641 \times 4 \times 4$. These volumes are
factored into the product of two $(641\times 4)\times 340$ matrices
where $340$ is the rank $r$. After reconstruction, the results of which
are plotted in Figure~\ref{OriginalData} for a single shot record, we
recover shot gathers sampled at $12.5$m from receivers collected at
random at an average receiver spacing of $50$m. While we are able to
recover this shot record, strong noisy artifacts remain especially at
the long offsets. In addition, important reflection and diffraction
information is missing and the ground roll is not well recovered making
this wavefield recovery unsuitable for subsequent processing.

\vspace*{-0.45cm}

\section{Proposed practical workflow}\label{proposed-practical-workflow}

\begin{figure}
\centering
\includegraphics[width=1.000\hsize]{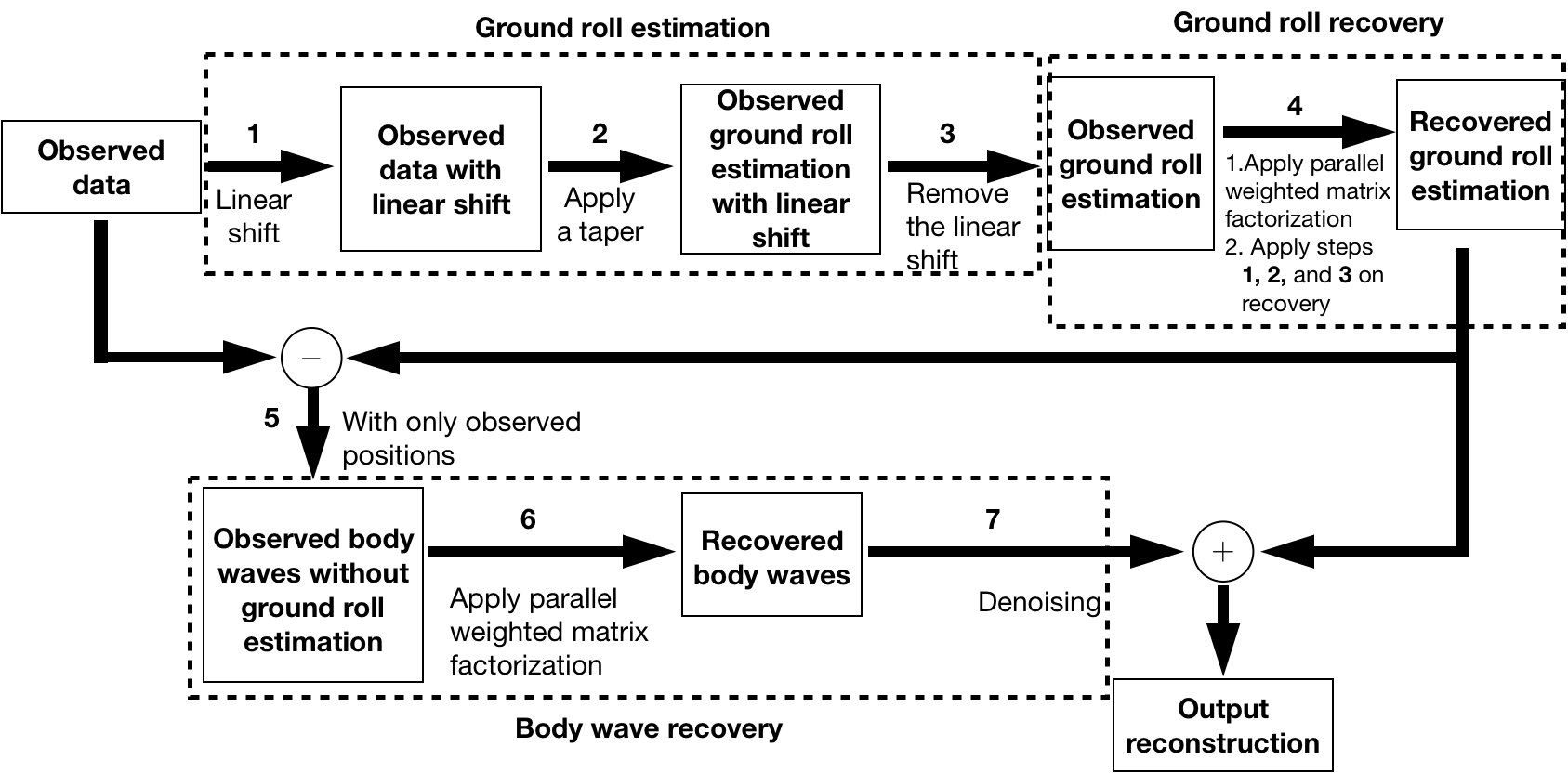}
\caption{Flowchart of proposed method.}\label{Workflow}
\end{figure}

To mitigate the effects of ground roll, we propose the reconstruction of
the body and surface (ground roll) waves separately. In this approach,
outlined in Figure~\ref{Workflow}, we use the fact that ground roll is
slow and relatively easily separable by applying a linear shift to the
data. Below, we describe the different steps outlined in the dashed
boxes in Figure~\ref{Workflow}.

\subsection{Ground roll estimation}\label{ground-roll-estimation}

\begin{figure}
\centering
\subfloat[\label{obs}]{\includegraphics[width=0.500\hsize]{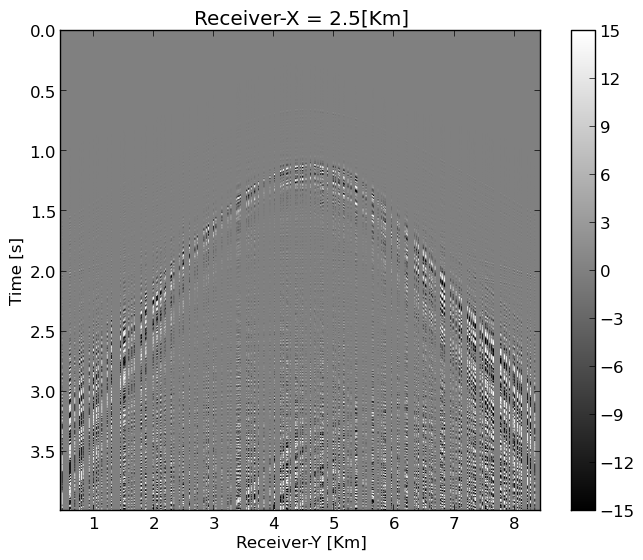}}
\subfloat[\label{S_obs}]{\includegraphics[width=0.500\hsize]{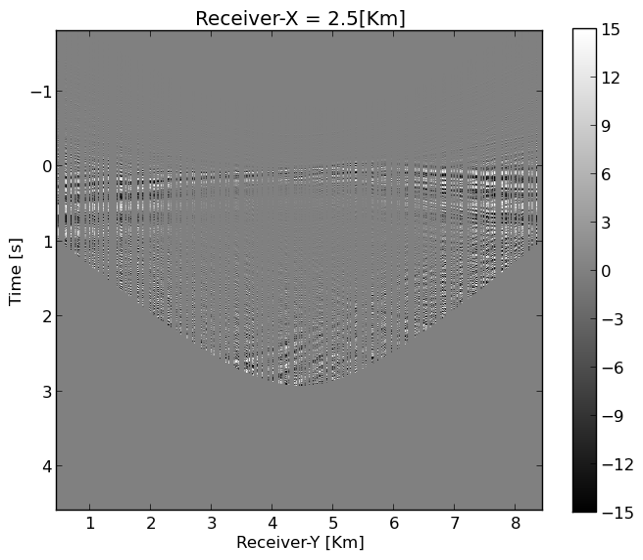}}
\\
\subfloat[\label{ST_obs}]{\includegraphics[width=0.500\hsize]{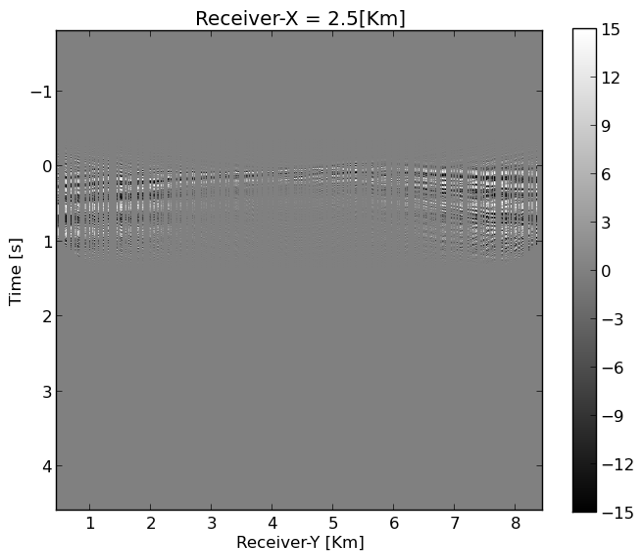}}
\subfloat[\label{RG_obs}]{\includegraphics[width=0.500\hsize]{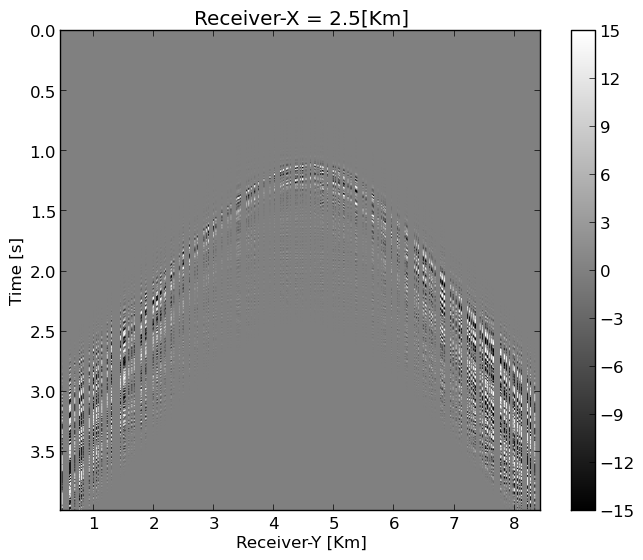}}
\caption{Ground roll estimation. \emph{(a)} Input shot record.
\emph{(b)} Input after linear shift. \emph{(c)} Tapered data. \emph{(d)}
Estimate subsampled ground roll after undoing linear
shift.}\label{ExtractGroundRoll1}
\end{figure}

Because the ground roll is slow, it is steep and therefore at least in
an approximate sense, separable from the body waves. This allows us to
devise a separate reconstruction scheme to recover the ground roll
before adding it back to the reconstruction of the body waves. We obtain
an estimate of the ground roll by carrying out the following steps: (1)
after zero-padding the input data (Figure~\ref{obs}), we apply a linear
shift aligning the ground roll (Figure~\ref{S_obs}); (2) we apply a
smooth taper with smooth cutoffs around at $t=0$s and $t=1$s designed to
extract the ground roll (Figure~\ref{ST_obs}), followed by (3) undoing
the linear shift, yielding an estimate of the randomly subsampled ground
roll plotted in Figure~\ref{RG_obs}. This estimate for the ground roll
serves as input for the reconstruction.

\subsection{Ground roll recovery}\label{ground-roll-recovery}

We use the estimated ground roll as input to our wavefield
reconstruction based on weighted matrix factorizations for a rank
$r=250$, which we find empirically by observing continuity of signals
and limited noise in the reconstructed data. We run the reconstruction
over all shots simultaneously for $320$ iterations of SPG-$\ell_2$
\citep{lopez2015rank} per frequency slice. To avoid reconstruction
leakage, we apply steps \emph{(1)-(3)} from the previous section again
to get the final ground roll recovery.

\subsection{Body wave recovery}\label{body-wave-recovery}

\begin{figure}
\centering
\subfloat[\label{obs_UG}]{\includegraphics[width=0.500\hsize]{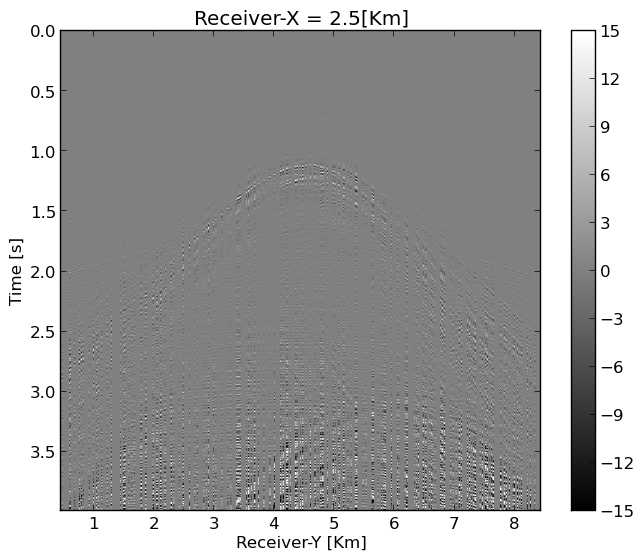}}
\subfloat[\label{Result1}]{\includegraphics[width=0.500\hsize]{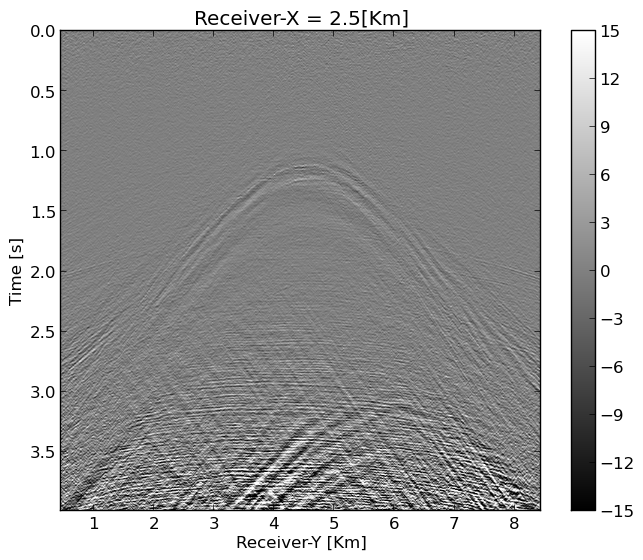}}
\caption{Body waves in the time domain. \emph{(a)} Observed body waves.
\emph{(b)} Reconstructed body waves}\label{Recon_reflective}
\end{figure}

\begin{figure}
\centering
\subfloat[\label{obs_F}]{\includegraphics[width=0.500\hsize]{results/Observed_RX.png}}
\subfloat[\label{Result_F}]{\includegraphics[width=0.500\hsize]{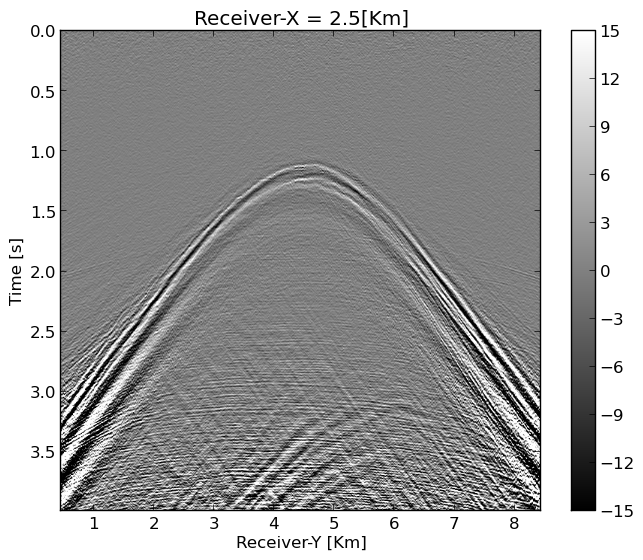}}
\caption{Seismic data reconstruction in the time domain. \emph{(a)}
Observed shot gather. \emph{(b)} Reconstructed shot
gather.}\label{FinalRecon}
\end{figure}

After reconstruction of the ground roll, we apply the mask
$\mathcal {A}$ to restrict the ground roll reconstruction to the
observed receiver positions again and subtract it from the original
subsampled input data. The resulting ``ground roll free'' estimate for
the body waves subsequently serves as input to a second wavefield
reconstruction now for the body waves. Since these waves are more
complex than ground roll, we choose the rank higher ($r=340$). As we can
observe from Figure~\ref{Recon_reflective}, the reconstructed body waves
contain, as expected, some remaining low-amplitude ground roll. Before
inverse Fourier transforming the reconstructed body waves, we apply a
$f-k$ filter to each shot along both receiver coordinates to remove
remnant noise. The resulting recovery for the body waves shown in
Figure~\ref{Result1} shows reconstruction of high frequency reflected
and diffracted energy. To arrive at the final result, we combine the
wavefield reconstructions for the ground roll and body waves. The result
of this blind study for a single shot is included in
Figure~\ref{FinalRecon}.

\vspace*{-0.45cm}

\section{Quality control (QC)}\label{quality-control-qc}

\begin{figure}
\centering
\subfloat[\label{G_truth2}]{\includegraphics[width=0.500\hsize]{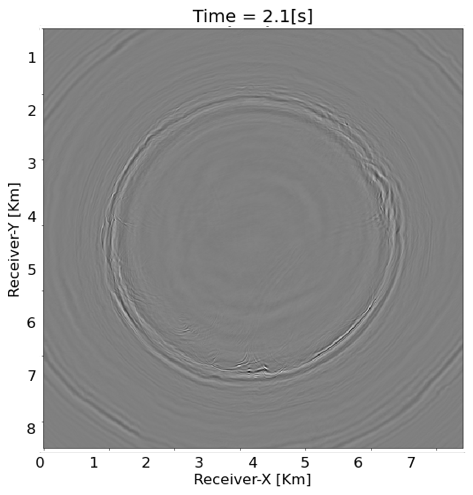}}
\subfloat[\label{Recon2}]{\includegraphics[width=0.500\hsize]{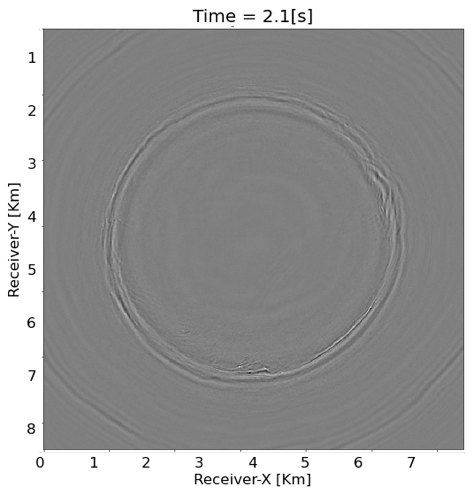}}
\\
\subfloat[\label{Diff2}]{\includegraphics[width=0.500\hsize]{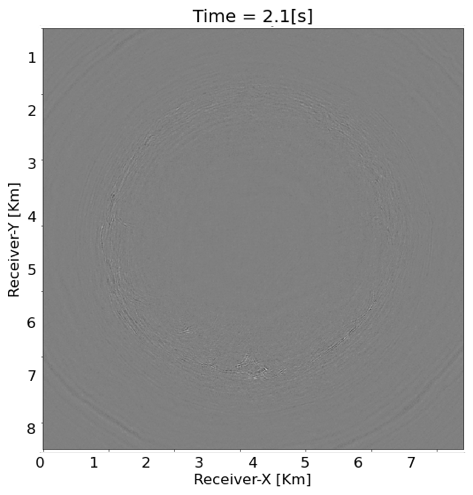}}
\caption{Wavefield recovery of one common shot gather. \emph{(a)} Time
slice at $2.1$s of ground truth. \emph{(b)} Time slice at $2.1$s of
reconstructed data. \emph{(c)} Time slice at $2.1$s of difference
between ground truth and recovery. All the subfigures are plotted on the
same scale.}\label{FinalPlots}
\end{figure}

By comparing Figure~\ref{Result_F} with Figure~\ref{UnweightedRX}, we
observe that the proposed method produces results with less artifacts at
the long offsets, and the ground roll is well recovered, especially at
the near offsets (see Figure~\ref{Result_F} for the receiver coordinate
$y$ between $4-6$ km for the time interval $1-1.5$ s).

To further verify our proposed practical workflow, we sent one
reconstructed shot gather to Occidental and obtained the following plots
in return. Figure~\ref{FinalPlots} contains time slices at $2.1$s with
the ground truth (Figure~\ref{G_truth2}), reconstructed wavefield
(Figure~\ref{Recon2}) and difference plot (Figure~\ref{Diff2}). From
these plots, we observe that the proposed method successfully recovers
body waves (reflections and diffractions) despite the presence of strong
aliased ground roll. During our presentation, we will further QC our
results by working with Occidental to produce stacks.

\vspace*{-0.45cm}

\section{Conclusions}\label{conclusions}

We presented a practical workflow successfully recovering a subset of
the synthetic $3$D SEAM Barrett dataset randomly sampled at $25$ percent
receiver sampling. Our workflow consists of the combination of a
weighted matrix factorization scheme and a separation of the subsampled
input data into ground roll and body wave components. Thanks to this
decomposition, we were able to mitigate the effects induced by the
strong aliased ground roll. Initial findings of the blind test we
carried out in collaboration with Occidental show that our method is
capable of dealing with ground roll while recovering high-frequency body
waves.

\vspace*{-0.45cm}

\section{Related materials}\label{related-materials}

The Julia code for this work is available on the
\href{https://slim.gatech.edu}{SLIM} GitHub page
\url{https://github.com/slimgroup/Software.SEG2021}.

\vspace*{-0.45cm}

\section{Acknowledgement}\label{acknowledgement}

This research was carried out with the support of Georgia Research
Alliance and partners of the ML4Seismic Center. We would like to
acknowledge the support from Occidental for providing the dataset, and
Georgia Institute of Technology for funding this research. We also would
like to thank Klaas Koster for assisting us to carry out the blind test.

\bibliography{abstract}

\begin{thebibliography}{}
\itemsep0pt

\bibitem[Aravkin et~al., 2014]{aravkin2014fast}
Aravkin, A., R. Kumar, H. Mansour, B. Recht, and F.~J. Herrmann,  2014, Fast
  methods for denoising matrix completion formulations, with applications to
  robust seismic data interpolation: SIAM Journal on Scientific Computing, {\bf
  36}, S237--S266.

\bibitem[Bahia et~al., 2020]{bahia2020ground}
Bahia, B., I. Papathanasaki, and M.~D. Sacchi,  2020, Ground-roll attenuation
  through quaternionic inversion with sparsity constraints, {\it in} SEG
  Technical Program Expanded Abstracts 2020: Society of Exploration
  Geophysicists,  3254--3258.

\bibitem[Cand{\`e}s et~al., 2006]{candes2006robust}
Cand{\`e}s, E.~J., J. Romberg, and T. Tao,  2006, Robust uncertainty
  principles: Exact signal reconstruction from highly incomplete frequency
  information: IEEE Transactions on information theory, {\bf 52}, 489--509.

\bibitem[Chiu, 2019]{chiu20193d}
Chiu, S.~K.,  2019, 3d attenuation of aliased ground roll on randomly
  undersampled data, {\it in} SEG Technical Program Expanded Abstracts 2019:
  Society of Exploration Geophysicists,  4560--4564.

\bibitem[Eftekhari et~al., 2018]{eftekhari2018weighted}
Eftekhari, A., D. Yang, and M.~B. Wakin,  2018, Weighted matrix completion and
  recovery with prior subspace information: IEEE Transactions on Information
  Theory, {\bf 64}, 4044--4071.

\bibitem[Herrmann and Hennenfent, 2008]{herrmann2008non}
Herrmann, F.~J., and G. Hennenfent,  2008, Non-parametric seismic data recovery
  with curvelet frames: Geophysical Journal International, {\bf 173}, 233--248.

\bibitem[Kumar et~al., 2015]{kumar2015efficient}
Kumar, R., C. Da~Silva, O. Akalin, A.~Y. Aravkin, H. Mansour, B. Recht, and
  F.~J. Herrmann,  2015, Efficient matrix completion for seismic data
  reconstruction: Geophysics, {\bf 80}, V97--V114.

\bibitem[Liu, 1999]{liu1999ground}
Liu, X.,  1999, Ground roll supression using the karhunen-loeve transform:
  Geophysics, {\bf 64}, 564--566.

\bibitem[Lopez et~al., 2015]{lopez2015rank}
Lopez, O., R. Kumar, and F.~J. Herrmann,  2015, Rank minimization via
  alternating optimization-seismic data interpolation: 77th EAGE Conference and
  Exhibition 2015, European Association of Geoscientists \& Engineers, 1--5.

\bibitem[Mosher et~al., 2014]{mosher2014increasing}
Mosher, C., C. Li, L. Morley, Y. Ji, F. Janiszewski, R. Olson, and J. Brewer,
  2014, Increasing the efficiency of seismic data acquisition via compressive
  sensing: The Leading Edge, {\bf 33}, 386--391.

\bibitem[Oropeza and Sacchi, 2011]{oropeza2011simultaneous}
Oropeza, V., and M. Sacchi,  2011, Simultaneous seismic data denoising and
  reconstruction via multichannel singular spectrum analysis: Geophysics, {\bf
  76}, V25--V32.

\bibitem[Sacchi et~al., 1998]{sacchi1998interpolation}
Sacchi, M.~D., T.~J. Ulrych, and C.~J. Walker,  1998, Interpolation and
  extrapolation using a high-resolution discrete fourier transform: IEEE
  Transactions on Signal Processing, {\bf 46}, 31--38.

\bibitem[Tan et~al., 2019]{tan2019seam}
Tan, J., T. Li, F. Jarrah, K. Lee, R. Holt, N.~V.~D. Coevering, and K. Koster,
  2019, Seam phase ii barrett model classic data study: Processing, imaging,
  and attributes analysis, {\it in} SEG Technical Program Expanded Abstracts
  2019: Society of Exploration Geophysicists,  3840--3844.

\bibitem[Trad et~al., 2003]{trad2003latest}
Trad, D., T. Ulrych, and M. Sacchi,  2003, Latest views of the sparse radon
  transform: Geophysics, {\bf 68}, 386--399.

\bibitem[Van De~Coevering et~al., 2019]{van2019sceptic}
Van De~Coevering, N., K. Koster, and R. Holt,  2019, A sceptic’s view of vvaz
  and avaz, {\it in} SEG Technical Program Expanded Abstracts 2019: Society of
  Exploration Geophysicists,  3230--3234.

\bibitem[Villasenor et~al., 1996]{villasenor1996seismic}
Villasenor, J.~D., R. Ergas, and P. Donoho,  1996, Seismic data compression
  using high-dimensional wavelet transforms: Proceedings of Data Compression
  Conference-DCC'96, IEEE, 396--405.

\bibitem[Yilmaz, 2001]{yilmaz2001seismic}
Yilmaz, {\"O}.,  2001, Seismic data analysis: Processing, inversion, and
  interpretation of seismic data: Society of exploration geophysicists.

\bibitem[Zhang et~al., 2019]{zhang2019high}
Zhang, Y., S. Sharan, and F.~J. Herrmann,  2019, High-frequency wavefield
  recovery with weighted matrix factorizations, {\it in} SEG Technical Program
  Expanded Abstracts 2019: Society of Exploration Geophysicists,  3959--3963.

\bibitem[Zhang et~al., 2020]{zhang2020wavefield}
Zhang, Y., S. Sharan, O. Lopez, and F.~J. Herrmann,  2020, Wavefield recovery
  with limited-subspace weighted matrix factorizations, {\it in} SEG Technical
  Program Expanded Abstracts 2020: Society of Exploration Geophysicists,
  2858--2862.

\end{thebibliography}

\end{document}